\documentclass[12pt, prd, showpacs]{revtex4}
\usepackage{amssymb}
\usepackage{amsmath}

\input{tcilatex}

\begin{document}

\title{Circular orbits and acceleration of particles by near-extremal dirty
rotating black holes: general approach}
\author{Oleg B. Zaslavskii}
\affiliation{Kharkov V.N. Karazin National University, 4 Svoboda Square, Kharkov, 61077,
Ukraine}
\email{zaslav@ukr.net}

\begin{abstract}
We study the effect of ultra-high energy particles collisions near the black
hole horizon (BSW effect) for two scenarios: when one of particle either (i)
moves on a circular orbit or (ii) plunges from it towards the horizon. It is
shown that such circular near-horizon orbits can exist for near-extremal
black holes only. This includes the innermost stable orbit (ISCO),
marginally bound orbit (MBO) and photon one (PhO). We consider generic
"dirty" rotating black holes not specifying the metric and show that the
energy in the centre of mass frame has the universal scaling dependence on
the surface gravity $\kappa $. Namely, $E_{c.m.}\sim \kappa ^{-n}$ where for
the ISCO $n=\frac{1}{3}$ in case (i) or $n=\frac{1}{2}$ in case (ii). For
the MBO and PhCO $n=\frac{1}{2}$ in both scenarios that agrees with recent
calculations of Harada and Kimura for the Kerr metric. We also generalize
the Grib and Pavlov's observations made for the Kerr metric. The magnitude
of the BSW effect on the location of collision has a somewhat paradoxical
character: it is decreasing when approaching the horizon.
\end{abstract}

\maketitle

rxi


\section{Introduction}

Recent discovery of the unbound growth of energies in collisions of
particles near the horizon of the Kerr black hole \cite{ban} provoked a
series of works in which this phenomenon was investigated in detail (see,
e.g. the recent works \cite{kerr} - \cite{fr} and references therein).
According to the tradition formed, I call this effect found by M. Ba\~{n}%
ados, J. Silk and S.M. West the BSW effect according to the names of its
authors. Irrespective of details, there are some general conditions for the
BSW effect to occur. Particles should collide in the vicinity of the horizon
in such a way that the special relation between the energy and angular
momentum of one of colliding particles should be satisfied. Both conditions
(proximity to the horizon and fine-tuning of the particle's parameters) are
naturally satisfied for the innermost stable circular orbit (ISCO) in the
near-extremal case. This kind of orbit was known to be important in
astrophysics of accretion disks around black holes \cite{72}, \cite{s}, \cite%
{bh}. Therefore, the high-energetic collisions near such orbits are
interesting not only from the theoretical viewpoint but can also have
astrophysical applications. The details of such collisions are investigated
in \cite{kerr}, \cite{spiral} for the Kerr metric.

Meanwhile, astrophysical black holes are surrounded by matter and are in
this sense "dirty". For them, the BSW effect also should occur since it is
the property inherent to generic rotating black holes \cite{prd}. Although
the full analysis requires the knowledge of the details of the space-time,
some essential features of this effect turn out to be model-independent that
can be thought of as the manifestation of universality of black hole physics
near the horizon. In the present paper we mainly consider two such issues:
(i) the dependence of the effect on the distance from the horizon and (ii)
the properties of ISCO and some other circular orbits - such as marginally
bound orbit (MBO) and the photon one (PhO) near the horizon for the
near-extremal black holes. Originally, the BSW effect was discussed for
maximally rotating black holes \cite{ban} that entailed some doubts with
respect to attempts to relate it to realistic astrophysical context \cite%
{bert}, \cite{ted}. Meanwhile, later on, it was shown that the scenario of
multiple scattering leads to the possibility of the BSW effect in the Kerr
background even for nonextremal black holes \cite{jlg} that was extended to
generic dirty black holes \cite{prd}.

In the present paper the accent is made on the properties of nonextremal
black holes but for the case when they become near-etremal, so the surface
gravity $\kappa \rightarrow 0$. It turns out in this situation the
dependence of the relevant quantities on $\kappa $ is universal and holds
for all models. In this way, we find the asymptotic expressions for the
characteristics of different types of circular orbits. Using these results,
we find the general dependence of the energy $E_{c.m.}$ of colliding
particles in the centre of mass frame on $\kappa $.

Exact values for the characteristics of the circular orbits found in \cite%
{72} for the Kerr metric are rather cumbersome and even the check of
corresponding formulas is not so simple. However, it turns out that if one
is interested in the orbits' characteristics near the horizon of a
near-extremal black \ hole, the behavior of these quantities is rather
simple and can be found in a general explicit form.

Quite recently, the BSW effect on the ISCO was also studied for static black
holes in a magnetic field with the discussion of potential astrophysical
relevance \cite{fr}. Although this is a quite different issue to which our
approach does not apply directly, this shows potential relevance of the BSW
effect in astrophysics and serves as an additional motivation to study
properties of this effect on the near-horizon circular orbits.

In what follows one should distinguish between two variants of the BSW\
effect. The first variant implies that a particle orbiting the ISCO (MBO or
PhO) collides with some other particle (we call it O - variant from
"orbiting"). The second variant implies that one of colliding particles
plunges towards the horizon from a circular orbit having the same values of
energy and angular momentum which it had there (we call it H - variant from
"horizon"). In both variants, collisions occur in the immediate vicinity of
the horizon but the scenarios of collisions are somewhat different. It is
worth noting that the possibility of unbound growth of the energy in the
centre of mass frame due to collision between a particle on the circular
orbit with a radially moving one was already briefly mentioned in literature
(see eq. 2.55 in \cite{pir}).

\section{General properties of BSW effect for nonextremal black holes}

\subsection{Basic formulas}

Consider a generic axially symmetric rotating black hole space-time.\ Its
metric can be written as%
\begin{equation}
ds^{2}=-N^{2}dt^{2}+g_{\phi \phi }(d\phi -\omega dt)^{2}+dl^{2}+g_{zz}dz^{2}.
\label{z}
\end{equation}%
Here, the metric coefficients do not depend on $t$ and $\phi $. On the
horizon $N=0$. Alternatively, one can use coordinates $\theta $ and $r$,
similar to the Boyer--Lindquist ones for the Kerr metric, instead of $l$ and 
$z$. In (\ref{z}) we assume that the metric coefficients are even functions
of $z$, so the equatorial plane $\theta =\frac{\pi }{2}$ ($z=0$) is a
symmetry one. Throughout the paper we assume that the fundamental constants $%
G=c=$%
h{\hskip-.2em}\llap{\protect\rule[1.1ex]{.325em}{.1ex}}{\hskip.2em}
$=1$.

In the space-time under discussion there are two conserved quantities $%
u_{0}\equiv -E$ and $u_{\phi }\equiv L$ where $u^{\mu }=\frac{dx^{\mu }}{%
d\tau }$ is the four-velocity of a test massive particle, $\tau $ is the
proper time (or the affine parameter if the particle is lightlike) and $%
x^{\mu }=(t,\phi ,l,z)$ are coordinates. The aforementioned conserved
quantities have the physical meaning of the energy per unit mass (or
frequency for a lightlike particle) and the azimuthal component of the
angular momentum, respectively. It follows from the symmetry reasonings that
there exist geodesics in such a background which lie entirely in the plane $%
\theta =\frac{\pi }{2}$. For them, the equations of motion read (dot denotes
the derivative with respect to the proper time $\tau $):%
\begin{equation}
\dot{t}=u^{0}=\frac{E-\omega L}{N^{2}}.  \label{t}
\end{equation}%
We assume that $\dot{t}>0$, so that $E-\omega L>0$ outside the horizon. On
the horizon itself, $E-\omega L=0$ is allowed.%
\begin{equation}
\dot{\phi}=\frac{L}{g_{\phi \phi }}+\frac{\omega (E-\omega L)}{N^{2}},
\label{phi}
\end{equation}%
\begin{equation}
N^{2}\dot{l}^{2}\equiv -V_{eff}=(E-\omega L)^{2}-bN^{2}\text{, }b=(\gamma +%
\frac{L^{2}}{g_{\phi \phi }})\text{.}  \label{n}
\end{equation}

Here, $\gamma =0$ for lightlike geodesics and $\gamma =1$ for timelike ones.
For definiteness, we consider a pair of particles having the equal masses $%
m_{1}=m_{2}=m$. We also assume that either both particles are approaching
the horizon or one of them stands at a fixed $l$ while the other particle
moves towards the horizon, so $\dot{l}\leqslant 0$. The particles are
labeled by subscript $i=1,2$.

The quantity which is relevant for us is the energy in the centre of mass
frame $E_{c.m.}=\sqrt{2}m\sqrt{1-u_{\mu (1)}u^{\mu (2)}}$ \cite{ban}. After
simple manipulations, one obtains from (\ref{t}) - (\ref{n}) that%
\begin{equation}
\frac{E_{c.m.}^{2}}{2m^{2}}=h+1-\frac{L_{1}L_{2}}{g_{\phi \phi }}\text{ }
\label{en}
\end{equation}%
where%
\begin{equation}
h=\frac{X_{1}X_{2}-Z_{1}Z_{2}}{N^{2}}\text{, }  \label{x}
\end{equation}%
\begin{equation}
\,X_{i}\equiv E_{i}-\omega L_{i},  \label{xi}
\end{equation}%
\begin{equation}
Z_{i}=\sqrt{\,\left( X_{i}\right) ^{2}-N^{2}b_{i}}\text{, }  \label{z1}
\end{equation}%
\begin{equation}
b_{i}=1+\frac{L_{i}^{2}}{g_{\phi \phi }}\text{.}  \label{b}
\end{equation}

By definition, we call a particle critical if $\left( X_{i}\right) _{+}=0$,
so 
\begin{equation}
E_{i}=\omega _{+}L_{i}  \label{cr}
\end{equation}%
and usual otherwise. Here, subscript "+" refers to the values on the horizon.

\subsection{Dependence of the BSW effect on proximity to the horizon}

It was observed in \cite{gpp} that, rather surprisingly, the magnitude of
the BSW effect in the Kerr background decreases as a point of collision is
chosen closer and closer to the horizon. Now, we will show that this
property has an universal character. We take particle 1 to be near-critical,
so $E_{1}=\omega _{+}L_{1}(1+\delta )$, $\delta \ll 1$. Particle 2 is
assumed to be usual. Near the horizon, $N\rightarrow 0$, $Z_{2}\approx
X_{2}\neq 0$, so 
\begin{equation}
\frac{E_{c.m.}^{2}}{2m^{2}}\approx \left( X_{2}\right) _{+}Y\text{, }Y=\frac{%
X_{1}-Z_{1}}{N^{2}}\text{, }
\end{equation}%
\begin{equation}
X_{1}=L(\omega _{+}-\omega )+\omega _{+}L_{+}\delta  \label{x1}
\end{equation}%
where $L_{+}=\frac{E}{\omega _{+}}$.

We assume that near the point of collision, $N^{2}\sim \delta ^{2}$ since it
is just the case when both terms inside the square root in (\ref{z1}) have
the same order (see below). Also, from now on, we assume that the black hole
is nonextremal and, similarly to the Kerr nonextremal metric, $\omega
-\omega _{+}\sim r-r_{+}\sim N^{2}$ where $r$ is the analog of the
Boyer-Lindquist coordinate. Then, the first term in the right hand side of (%
\ref{x1}) has the order $\delta ^{2}$ and is negligible, so 
\begin{equation}
X_{1}\approx \omega _{+}L_{+}\delta \text{.}
\end{equation}%
It is convenient to make an ansatz introducing the new function $\chi $
according to%
\begin{equation}
N=\omega _{+}L_{+}\frac{\delta }{\sqrt{\left( b_{1}\right) _{+}}}\chi \text{.%
}
\end{equation}%
Then,%
\begin{equation}
Z_{1}\approx \omega _{+}L_{+}\delta \sqrt{1-\chi ^{2}}\text{,}
\end{equation}%
\begin{equation}
Y\approx \frac{b_{1}}{\omega _{+}L_{+}\delta }f\text{,}
\end{equation}%
where

\begin{equation}
f=\frac{1-\sqrt{1-\chi ^{2}}}{\chi ^{2}}.
\end{equation}

Thus we have an universal coordinate dependence in terms of the function $f$%
. It is convenient to make the substitution%
\begin{equation}
\chi =\sin \eta \text{, \thinspace }0\leq \eta \leq \frac{\pi }{2}\text{.}
\end{equation}%
Then,%
\begin{equation}
f=\frac{1}{2\cos ^{2}\frac{\eta }{2}}.
\end{equation}%
In the near-horizon region,

\begin{equation}
b=b_{+}+O(r-r_{+})=b_{+}+O(N^{2})=b_{+}+O(\delta ^{2})\text{.}
\end{equation}%
Thus neglecting the terms of the order $\delta ^{2}$, we have%
\begin{equation}
\frac{E_{c.m.}^{2}}{2m^{2}}\approx \left( X_{2}\right) _{+}\frac{b_{+}}{%
\omega _{+}L_{+}\delta }f\left( \chi \right) \text{.}
\end{equation}%
We obtained a function that (for a fixed small value of $\delta $) is
monotonically decreasing away from the horizon where $\eta =0$, $f=\frac{1}{2%
}$, to the turning point of particle 1 where $Z_{1}=0$ and $\eta =\frac{\pi 
}{2}$, $f=1$. In this sense, the dependence of the effect on the distance
has somewhat paradoxical character: the closer the point of collision is to
the horizon, the weaker its magnitude measured by $f$. This generalizes the
corresponding observation made for the Kerr metric in \cite{gpp}.

\section{Circular orbits}

We restrict ourselves to orbits in the equatorial plane. It is the
properties of circular orbits which we now turn to. It is convenient to
introduce, instead of the proper distance $l$, the radial coordinate $\rho $
according to $dl=\frac{d\rho }{N}$ . By definition, the circular orbit at $%
\rho =\rho _{0}$ is determined by the equalities%
\begin{equation}
V_{eff}(\rho _{0})=0\text{,}  \label{v}
\end{equation}%
\begin{equation}
\frac{dV_{eff}}{d\rho }(\rho _{0})=0  \label{v'}
\end{equation}%
where according to (\ref{n}) and (\ref{z1}), 
\begin{equation}
V_{eff}=-Z^{2}.  \label{vz}
\end{equation}%
These equations state that the $\rho _{0}$ is a perpetual turning point. It
follows from (\ref{n}), (\ref{z1}), (\ref{v}) that for a particle on such an
orbit, 
\begin{equation}
Z=0\text{, }X=\sqrt{b}N\text{.}  \label{zxi}
\end{equation}%
If this orbit lies near the horizon (see below for details), it follows from
(\ref{phi}) that 
\begin{equation}
\dot{\phi}\approx \frac{\omega _{+}\sqrt{b_{+}}}{N}>0
\end{equation}%
assuming $\omega _{+}>0$. Thus near the horizon a particle rotates on the
prograde orbit.

\subsection{Nonexistence of near-horizon circular orbits for generic
nonextremal rotating black holes}

First of all, we show that for a nonextremal black hole with a fixed surface
gravity $\kappa \neq 0$, there are no circular orbits in the near-horizon
limit, thus generalizing the observation of \cite{gpp} made there for the
Kerr metric. It follows from (\ref{n}) that%
\begin{equation}
-\frac{1}{2}\frac{dV_{eff}}{d\rho }=[-(E-\omega L)L\frac{d\omega }{d\rho }-%
\frac{dN}{dl}b-\frac{N^{2}}{2}\frac{db}{d\rho }]\text{.}
\end{equation}%
In the horizon limit $N\rightarrow 0$, $\frac{dN}{dl}\rightarrow \kappa .$
It follows from (\ref{n}), (\ref{v}) that in this limit 
\begin{equation}
X_{+}=E-\omega _{+}L\rightarrow 0\text{.}
\end{equation}%
We obtain that%
\begin{equation}
-\frac{1}{2}\frac{dV_{eff}}{d\rho }\rightarrow -b_{+}\kappa \neq 0
\end{equation}%
in contradiction with (\ref{v'}) that proves the statement.

In what follows, we will consider a near-extremal black hole, with small but
nonzero $\kappa $. When $\kappa $ itself tends to zero, the aforementioned
general prohibition does not work and circular orbits in the near-horizon
region can exist. Now, we will examine different kinds of orbits separately.

\subsection{Innermost stable circular orbit (ISCO)}

This is the circular orbit closest to the horizon on the threshold of the
stability. Correspondingly, we must add to equations (\ref{v}) and (\ref{v'}%
) also%
\begin{equation}
\frac{d^{2}V_{eff}(\rho _{0})}{d\rho ^{2}}=0\text{.}  \label{v''}
\end{equation}%
Now, we will find explicit formulas for the metric and dynamic
characteristics of a particle on the ISCO near the horizon of the nearly
extremal rotating black hole. We can use the power-expansion in terms of $%
x=\rho _{0}-\rho _{+}$ near the horizon $\rho =\rho _{+}$:%
\begin{equation}
N^{2}=2\kappa x+Dx^{2}+Cx^{3}...\text{,}  \label{N}
\end{equation}%
\begin{equation}
\omega =\omega _{+}-B_{1}x+B_{2}x^{2}...\text{ },  \label{om}
\end{equation}%
\begin{equation}
L=L_{0}+ax+...\text{.}  \label{L}
\end{equation}

Let me remind that the coefficients entering expansion (\ref{N}) are related
to the fixed coordinate gauge in which the coefficient at $d\rho ^{2}$ is
equal to $N^{-1}\,.$

It is instructive to stress that an orbit for which calculations are being
carried out lies outside the horizon. Moreover, although formally $\frac{%
\rho -\rho _{+}}{\rho _{+}}\rightarrow 0$ when $\kappa \rightarrow 0$ the
proper distance from the horizon does not vanish and even may be large (see
Sec. IV below for details). Therefore, $L_{0}$ is not the value of the
angular momentum on the horizon (where a massive particle cannot be situated
at all). Rather, it gives the value of the momentum on the near-horizon
orbit under discussion in the main approximation while the terms of the
order $x$ and higher give corrections to it. In doing so, $x$ is small but
cannot vanish on the ISCO, out goal is just to find it (see eq. (\ref{isco})
below and explicit comparison with the Kerr case in eq. (\ref{l1})).

We have two small values: $x$ and $\kappa $. We are interested in the
near-extremal limit. By definition, its very meaning consists in that in (%
\ref{N}) the first term is small as compared to the second one:

\begin{equation}
\kappa \ll Dx\text{.}
\end{equation}%
Correspondingly, in the region under discussion,%
\begin{equation}
N=x\sqrt{D}+\frac{\kappa }{\sqrt{D}}-\frac{\kappa ^{2}}{2D^{3/2}x}+\frac{C}{2%
\sqrt{D}}x^{2}+...  \label{nx}
\end{equation}

We substitute aforementioned expansions into (\ref{v}), (\ref{v'}) where (%
\ref{vz}), (\ref{z1}) are used. Then, in this limit, calculating $\frac{%
dV_{eff}}{d\rho }$ and neglecting high-order corrections we have 
\begin{equation}
-\frac{1}{2}\frac{dV_{eff}}{d\rho }(\rho _{0})\approx A_{0}\kappa
+A_{1}x+A_{2}x^{2}+A_{3}\frac{\kappa ^{2}}{x}=0  \label{vv}
\end{equation}%
with coefficients which are given below. Eq. (\ref{v''}) with (\ref{v}), (%
\ref{zxi}) taken into account gives us%
\begin{equation}
-\frac{1}{2}\frac{d^{2}V}{d\rho ^{2}}(\rho
_{0})=B_{1}^{2}L^{2}-bD-Fx+O(x^{2},\kappa x)=0\text{.}  \label{vf}
\end{equation}%
where%
\begin{equation}
F=4B_{1}B_{2}L^{2}+2L\sqrt{b_{0}}B_{2}\sqrt{D}+2Db^{\prime }+3bC\text{.}
\label{F}
\end{equation}%
We must substitute into (\ref{vv})\ the relationships that follow from eqs. (%
\ref{v}), (\ref{vf}), (\ref{F}). After some algebraic manipulations, one
finds:

\begin{equation}
A_{0}=A_{1}=0\text{, }
\end{equation}%
\begin{equation}
A_{2}=b_{0}D\frac{B_{2}}{B_{1}}+\frac{b_{0}^{\prime }D}{2}-b_{0}C\text{,}
\end{equation}%
\begin{equation}
A_{3}=-\frac{b_{0}}{2D}\text{,}
\end{equation}%
$b_{0}\equiv b(\rho _{0})$. Then, it follows from eq. (\ref{vv}) that

\begin{equation}
x\approx H\kappa ^{2^{\prime }3}\text{, }H=(\frac{b_{0}}{2DA_{2}})^{1/3}%
\text{.}  \label{isco}
\end{equation}

To express $L_{0}$ entirely in terms of the metric coefficients on the
horizon, one should take into account the definition of $b$ in eq. (\ref{b})
and eq. (\ref{vf}). As a result, we have%
\begin{equation}
L\approx L_{0}+\frac{FH}{2\sqrt{D(B_{1}^{2}-\frac{D}{g_{\phi \phi }})}}%
\kappa ^{\frac{2}{3}}\text{,}  \label{l01}
\end{equation}%
\begin{equation}
L_{0}^{2}=\frac{D}{B_{1}^{2}-\frac{D}{g_{\phi \phi }}}\text{, }b_{+}=\frac{%
B_{1}^{2}\left( g_{\phi \phi }\right) _{+}}{B_{1}^{2}\left( g_{\phi \phi
}\right) _{+}-D}\text{.}  \label{l0}
\end{equation}

Eqs. (\ref{v}), (\ref{nx}) and (\ref{isco}) give us%
\begin{equation}
X=\sqrt{b_{0}}N\approx \sqrt{b_{0}}\sqrt{D}H\kappa ^{\frac{2}{3}}\text{,}
\label{X}
\end{equation}

so on the ISCO 
\begin{equation}
X\sim N\sim \kappa ^{\frac{2}{3}}.  \label{xn}
\end{equation}

Thus the quantity $X$ that according to (\ref{xi}), (\ref{cr}) measures the
deviation of particle's parameters from the criticality has an universal
dependence on the surface gravity irrespective of the concrete form of the
metric.

\subsection{Marginally bound orbit (MBO)}

Another type of circular orbit is a so-called marginally bound one (MBO).
This means that a particle satisfying eqs. (\ref{v}) and (\ref{v'}) has the
zero velocity at infinity, so the energy per mass $E=1$. We are interested
in the prograde MBO only since for it $\rho _{0}\rightarrow \rho _{+}$ in
the extremal limit. After some algebra, one finds from (\ref{n}), (\ref{v}),
(\ref{v'}), (\ref{N}) and (\ref{om}) that now for small $\kappa $, 
\begin{equation}
x\approx \kappa \alpha
\end{equation}%
where the coefficient $\alpha $ is \ equal to%
\begin{equation}
\alpha =\frac{1}{D}(\frac{1}{\sqrt{1-\frac{D}{P}}}-1)\text{,}  \label{al}
\end{equation}%
\begin{equation}
\frac{1}{P}\equiv \frac{1}{B_{1}^{2}}[\frac{1}{\left( g_{\phi \phi }\right)
_{+}}+\omega _{+}^{2}]=\frac{\omega _{+}^{2}}{B_{1}^{2}}b_{+}\text{,}
\label{bm}
\end{equation}%
\begin{equation}
E=\omega _{+}+\frac{\kappa B_{1}}{\omega _{+}\sqrt{P}\sqrt{P-D}}>0\text{,}
\label{eh}
\end{equation}%
\begin{equation}
N^{2}(\rho _{0})\approx \kappa ^{2}\frac{1}{P-D}\text{.}  \label{ND}
\end{equation}%
\begin{equation}
L=\frac{1}{\omega _{+}}-\kappa s\text{, }s=\frac{B_{1}}{D\omega _{+}^{2}}(1-%
\sqrt{1-\frac{D}{P}})>0\text{,}  \label{lm}
\end{equation}%
\begin{equation}
b_{+}=1+\frac{1}{\left( g_{\phi \phi }\right) _{+}\omega _{+}^{2}}=\frac{%
B_{1}^{2}}{\omega _{+}^{2}P}\text{.}  \label{b1}
\end{equation}

In this case, eq. (\ref{zxi}) gives us 
\begin{equation}
X\approx \sqrt{b_{+}}N\approx \frac{1}{\omega _{+}}\frac{B_{1}}{\sqrt{P}}%
\frac{\kappa }{\sqrt{P-D}}\text{,}  \label{xnb}
\end{equation}%
so%
\begin{equation}
X\sim N\sim \kappa \text{.}  \label{mb}
\end{equation}

\subsection{Photon orbits (PhO)}

Now, for massless particles (photons) one must put $\gamma =0$ in (\ref{n}).
Then, we have for the reduced potential $\tilde{V}\equiv \frac{V_{eff}}{L^{2}%
}$:

\begin{equation}
-\tilde{V}=(\tilde{E}-\omega )^{2}-\frac{N^{2}}{g_{\phi \phi }}\text{, }%
\tilde{E}=\frac{E}{L}.
\end{equation}%
For the photon circular orbit, $\tilde{V}(\rho _{0})=\tilde{V}^{\prime
}(\rho _{0})=0$, whence at $\rho =\rho _{0}$ we have%
\begin{equation}
(\tilde{E}-\omega )^{2}-\frac{N^{2}}{g_{\phi \phi }}=0\text{,}  \label{ph1}
\end{equation}%
\begin{equation}
2(\tilde{E}-\omega )\frac{d\omega }{d\rho }+\left( \frac{N^{2}}{g_{\phi \phi
}}\right) ^{\prime }=0.  \label{ph2}
\end{equation}%
One can verify that for the Schwarzschild case one obtains form here the
well-known result $\rho _{0}=3M$.

For a near-extremal black hole, in the near-horizon region, it is easy to
check that in expansion (\ref{N}) the first and second terms have the same
order, $x=\kappa \alpha $ where in the main approximation $\alpha $ does not
contain $\kappa $. Collecting in eqs. (\ref{ph1}) and (\ref{ph2}) all terms
of the first order in $\kappa $ and solving corresponding equation with
respect to $\alpha $, one obtains that eqs. (\ref{al}), (\ref{ND}) are now
satisfied with 
\begin{equation}
P=B_{1}^{2}\left( g_{\phi \phi }\right) _{+}\text{.}
\end{equation}

Then, we obtain for $\tilde{E}$ the general expression 
\begin{equation}
\tilde{E}=\omega _{+}+\kappa \frac{B_{1}}{D}(1-\frac{\sqrt{P-D}}{P})>0\text{.%
}
\end{equation}%
\begin{equation}
X\approx \frac{L}{\sqrt{\left( g_{\phi \phi }\right) _{+}}}\frac{\kappa }{%
\sqrt{P-D}}=\frac{B_{1}L}{\sqrt{P}}\frac{\kappa }{\sqrt{P-D}}\text{.}
\end{equation}%
Now, the dependence (\ref{mb}) still holds.

\section{Proper distance.}

All three kinds of orbits have $\rho \approx \rho _{+}$. However, they are
separated spatially since the proper distance between them does not vanish.
This generalizes the corresponding properties of the Kerr metric \cite{72}.
Namely, between the horizon and the MBO or PhO, writing $\rho -\rho
_{+}=\kappa y$, we have 
\begin{equation}
l=\int \frac{d\rho }{N}\approx \int_{0}^{\alpha }\frac{dy}{\sqrt{2y+Dy^{2}}}=%
\frac{1}{\sqrt{D}}\ln \frac{\sqrt{P}+\sqrt{D}}{\sqrt{P-D}}  \label{l}
\end{equation}%
where eq. (\ref{al}) was taken into account.

Between the horizon and the ISCO, 
\begin{equation}
l\approx \frac{1}{\sqrt{D}}\ln \frac{(\rho _{0}-\rho _{+})}{\kappa }\approx 
\frac{1}{3\sqrt{D}}\ln \frac{1}{\kappa }\text{.}
\end{equation}

\section{Near-extremal Kerr}

It is instructive to check these formulas for the Kerr case. Then,

\begin{equation}
\frac{d\rho ^{2}}{N^{2}}=r^{2}\frac{dr^{2}}{\Delta }\text{, }\Delta
=r^{2}-2Mr+a^{2}=(r-r_{+})(r-r_{-})
\end{equation}%
where $r$ is the Boyer--Lindquist coordinate, $r_{\pm }=M\pm \sqrt{%
M^{2}-a^{2}}$, $M$ is the black hole mass, $a$ characterizes its angular
momentum. Then, 
\begin{equation}
N^{2}=\frac{(r-r_{+})(r-r_{-})}{r^{2}+a^{2}+\frac{2Ma^{2}}{r}}\text{, }%
g_{\phi \phi }=r^{2}+a^{2}+\frac{2Ma^{2}}{r}\text{,}
\end{equation}%
\begin{equation}
\frac{d\rho }{dr}=\frac{r}{\sqrt{r^{2}+a^{2}+\frac{2Ma^{2}}{r}}},  \label{rr}
\end{equation}%
\begin{equation}
\omega =\frac{2Ma}{r^{3}+a^{2}r+2Ma^{2}}\text{, }\omega _{+}=\frac{a}{2Mr_{+}%
}  \label{w}
\end{equation}%
\begin{equation}
\kappa =\frac{\sqrt{M^{2}-a^{2}}}{2Mr_{+}}.
\end{equation}

In the near-extremal case 
\begin{equation}
a=M(1-\varepsilon ),  \label{a}
\end{equation}%
$\varepsilon \ll 1$, 
\begin{equation}
\kappa \approx \frac{\sqrt{\varepsilon }}{\sqrt{2}M}\text{,}  \label{k}
\end{equation}%
the horizon radius 
\begin{equation}
r_{+}\approx M(1+\sqrt{2\varepsilon })\text{,}  \label{rm}
\end{equation}%
the horizon angular velocity%
\begin{equation}
\omega _{+}\approx \frac{1}{2M}(1-\sqrt{2\varepsilon }).  \label{0mk}
\end{equation}

The relevant coefficients defined above are in the main approximation equal
to%
\begin{equation}
C=0\text{, }D=\frac{1}{M^{2}}\text{, }B_{1}=\frac{1}{M^{2}}\text{, }B_{2}=%
\frac{1}{2M^{3}}\text{, }H=M^{5/3}\text{, }F=\frac{4}{M^{3}}\text{.}
\label{cd}
\end{equation}

Now, we can obtain the radius $r_{0}$ and other main characteristics of all
circular orbits near the horizon.

\subsubsection{ISCO}

From (\ref{isco}) it follows that, neglecting high-order corrections,%
\begin{equation}
b_{0}=\frac{4}{3}\text{, }b_{0}^{\prime }=0\text{, }A_{2}=\frac{2}{3M^{3}}%
\text{, }H^{3}=M^{5}  \label{is}
\end{equation}%
\begin{equation}
\frac{r_{0}-M}{M}\approx 2^{2/3}\varepsilon ^{1/3},  \label{ris}
\end{equation}%
\begin{equation}
l\approx \frac{M}{6}\ln \frac{1}{\varepsilon }\text{,}
\end{equation}%
\begin{equation}
L\approx \frac{2}{\sqrt{3}}+\frac{2}{\sqrt{3}}2^{2/3}\varepsilon ^{1/3}\text{%
,}  \label{l1}
\end{equation}%
\begin{equation}
X\approx \frac{2^{2/3}}{\sqrt{3}}\varepsilon ^{1/3}.  \label{xx}
\end{equation}

\subsubsection{MBO}

Taking into account eqs. (\ref{k}), (\ref{rm}), (\ref{rr}), (\ref{w}), (\ref%
{0mk}) one finds that 
\begin{equation}
P=\frac{2}{M^{2}},\alpha =M^{2}(\sqrt{2}-1)\text{, }s=\frac{\sqrt{2}-1}{%
\sqrt{2}}  \label{d0}
\end{equation}

\begin{equation}
r-M\approx 2M\sqrt{\varepsilon }\text{, }
\end{equation}%
\begin{equation}
\frac{L}{M}\approx 2+2\sqrt{\varepsilon }\text{, }  \label{lmb}
\end{equation}%
\begin{equation}
E=1\text{, }\frac{L}{E}\approx 2M\text{.}
\end{equation}

The proper distance from the horizon%
\begin{equation}
l\approx M\ln (1+\sqrt{2}).
\end{equation}

\subsubsection{PhO}

Now, $P=\frac{4}{M^{2}}$, $\alpha =M(\frac{2}{\sqrt{3}}-1)$ and, taking into
account eq. (\ref{rm}), we obtain 
\begin{equation}
r_{0}-M=2M\sqrt{\frac{2}{3}}\sqrt{\varepsilon }\text{,}
\end{equation}%
\begin{equation}
\tilde{E}=\frac{E}{L}=\frac{1}{2M}\text{.}
\end{equation}

In the limit under discussion, the proper distance between the PhO and MBO
equals in this limit%
\begin{equation}
l=M\ln \frac{1+\sqrt{2}}{\sqrt{3}}\text{.}  \label{p}
\end{equation}

Eqs. (\ref{ris}) - (\ref{p}) agree with eqs. (2.22), (2.23) of Ref. \cite{72}%
. Eq. (\ref{xx}) agrees with eqs. (4.6), (4.7) of Ref. \cite{kerr}, eq. (\ref%
{l1}) reproduces the foirst two terms of eq. (4.7) of \cite{kerr}.

\section{O-variant of BSW effect: collisions on circular orbits}

Now, we apply the results for the characteristics of circular orbits to the
collisions and elucidate properties of the BSW effect for generic
near-extremal black holes. As for any circular orbits $Z=0$, in the
near-horizon limit this entails that $X\rightarrow 0$ according to (\ref{z1}%
) and (\ref{v}). The quantity $h$ that enters the expression for the energy (%
\ref{en}) simplifies:%
\begin{equation}
h=\frac{X_{1}X_{2}}{N^{2}}\text{.}  \label{cn2}
\end{equation}

It is worth noting that the dependence of the energy on parameters of two
particles has factorized. Now, from the results of previous sections, we
have already all means to find the asymptotic behavior of $h$ for small $%
\kappa $. When $h\rightarrow \infty $, for collision of two massive
particles $E_{c.m.}$ $\approx m\sqrt{2h}$ according to eq. (\ref{en}).

Further, we will consider different kinds of circular orbits separately.

\subsection{ISCO}

\subsubsection{Both particles move on ISCO}

Then, it follows from (\ref{xn}) that the quantity $h$ is finite and so is
the energy $E_{c.m.}$. Thus the BSW effect is absent. This can be easily
understood from the previous general results. Namely, a particle moving on
ISCO\ is necessarily near-critical since $X\rightarrow 0$ for it. Meanwhile,
collision between two critical particles cannot produce the BSW effect \cite%
{prd}, \cite{k}.

\subsubsection{One particle is on ISCO, second particle is usual}

Let particle 1 be on ISCO and particle 2 be usual. Then, it follows from (%
\ref{xn}) that $X_{1}\sim \kappa ^{2/3}\sim N$, so (\ref{en}) gives us

\begin{equation}
E_{c.m.}\approx Um\sqrt{2X_{2}}\kappa ^{-1/3}\sim \kappa ^{-1/3}\text{.}
\label{eu}
\end{equation}%
Here, the coefficient $U$ can be found from (\ref{nx}), (\ref{isco}), (\ref%
{X}):%
\begin{equation}
U=\left( \frac{b_{0}}{D}\right) ^{1/12}(2A_{2})^{1/6}.  \label{u}
\end{equation}

\subsection{MBO}

\subsubsection{Both particles on the MBO}

It is seen from (\ref{mb}) that $X_{1}\sim X_{2}\sim \kappa ^{1/2}\sim $ $N$%
. Therefore, $c$ is finite and the BSW effect is absent. Explanation is
similar to that for the ISCO.

\subsubsection{One particle is on the MBO, second particle is usual}

Now, we have from (\ref{xnb}), (\ref{b1}) and (\ref{en}) that%
\begin{equation}
E_{c.m.}\approx Vm\sqrt{2X_{2}}\kappa ^{-1/2}\sim \kappa ^{-1/2},
\label{cso}
\end{equation}%
\begin{equation}
V=\sqrt{\frac{B_{1}}{\omega _{+}}}(1-\frac{D}{P})^{1/4}\text{.}
\end{equation}

\subsection{PhO}

We assume that the second particle is massive with the mass $m$, so we are
dealing with the collision between massive and massless particles \cite{k}.
In the present situation, a photon is already critical. For collisions
between such particles, the formula for the energy $E_{c.m.\text{ }}$
preserves its general structure but now the expression for $b$ changes for
photon that affects also the expression for $Z$. Now, in eq. (\ref{z1}), $b=%
\frac{L^{2}}{g_{\phi \phi }}$ (see Sec. VI of Ref. \cite{prd} for details).
The situation does not differ qualitatively from that for MBO.

\subsubsection{Massive particle is critical.}

Using eq. (\ref{mb}), we see that the energy turns out to be finite in
accordance with the general conclusion of \cite{k} that two critical
particles are unable to produce the BSW effect.

\subsubsection{Massive particle is usual}

Now, according to \cite{k}, eq. (\ref{en}) is somewhat changed: 
\begin{equation}
\frac{E_{c.m.}^{2}}{2m^{2}}=\frac{1}{2}+\frac{1}{m}(h-\frac{L_{1}L_{2}}{%
g_{\phi \phi }})
\end{equation}%
where $h$ is given by the same expression (\ref{x}). Then, the dependence of
the energy $E_{c.m.}$ on $\kappa $ in the limit under discussion is the same
as in (\ref{cso}) although with another coefficients:%
\begin{equation}
E_{c.m.}\approx W\sqrt{2mX_{2}}\kappa ^{-1/2}\sim \kappa ^{-1/2}\text{,}
\end{equation}%
\begin{equation}
W=\sqrt{B_{1}L}(1-\frac{D}{P})^{1/4}\text{.}
\end{equation}

One should bear in mind that for massive particles $L$ has dimension of $M$,
while for massless ones $L$ does not contain $M$ since it enters the
combination $\nu -\omega L$, $\nu $ is the photon frequency.

\section{H-variant of BSW effect: collisions of particles plunging from
circular orbits}

In the previous section we dealt with the situation when a particle moving
along circular near-hoizon orbit collided with some usual particle (O -
variant of the BSW effect, according to our definition). Now, we consider
somewhat another scenario: both particles collide on the horizon, one of
them arrived there from a circular orbit (H-variant of the BSW effect,
according to our definition). It means that it spent some time on that orbit
and, due to instability of the orbit or being on the threshold of stability,
moved towards the horizon with the same values of the energy and momentum
which it had at the circular orbit. Mathematically, it corresponds to taking
the horizon limit $N\rightarrow 0$ first in the formula (\ref{en}) for $%
E_{c.m.}$. Then, one can derive the general expression which can be also
take directly from eq. (9) of \cite{prd}:%
\begin{equation}
\frac{E_{c.m.}}{m}\approx \sqrt{\frac{b_{+}Y_{2}}{Y_{1}}}  \label{e2}
\end{equation}

Here, 
\begin{equation}
Y_{i}=E_{i}-\omega _{+}L_{i}.
\end{equation}

The quantities $Y_{i}$ differ slightly from $X_{i}$ defined in (\ref{xi}):%
\begin{equation}
Y_{i}=X_{i}+(\omega -\omega _{+})L_{i}.  \label{yx}
\end{equation}

For a usual particle 2, $X_{2}\neq 0$, so we can safely neglect the second
small term in the right hand side of (\ref{yx}), whence $Y_{2}\approx X_{2}$%
. For near-critical particle 1 with small $X_{1}$, the situation is much
more subtle and one should be careful keeping both terms. As usual, we
consider different kinds of circular orbits separately.

\subsection{ISCO}

It follows from (\ref{zxi}), (\ref{nx}), (\ref{om}) that

\begin{equation}
Y_{1}=\sqrt{b}N+(\omega -\omega _{+})L_{1}=(\sqrt{b}\sqrt{D}-L_{1})x_{0}
\label{y1}
\end{equation}%
where $L_{0}$ is the value of the angular momentum on the ISCO. Meanwhile,
it follows from (\ref{l01}), (\ref{l0}) that in the main approximation $%
\sqrt{b}\sqrt{D}-L_{1}=0$. Thus both terms in $Y_{1}$ mutually cancel. The
main nonvanishing correction comes from the second term in (\ref{nx}).\
Then, we have%
\begin{equation}
Y_{1}\approx \sqrt{\frac{b}{D}}\kappa \text{.}
\end{equation}%
By substitution into (\ref{e2}), we obtain that%
\begin{equation}
E_{c.m.}\approx m\sqrt{b_{1}\frac{Y_{2}}{Y_{1}}}\approx m\sqrt{X_{2}}%
(bD)^{1/4}\kappa ^{-1/2}\text{.}  \label{is2}
\end{equation}

\subsection{MBO}

Now, it follows from (\ref{bm}) - (\ref{lm}) and (\ref{yx}) that 
\begin{equation}
Y_{1}=\kappa s\omega _{+}\text{,}
\end{equation}%
\begin{equation}
\frac{E_{c.m.}}{m}=A_{mbo}\sqrt{X_{2}}\kappa ^{-1/2}  \label{ambo}
\end{equation}%
\begin{equation}
A_{mbo}=\sqrt{\frac{DB_{1}}{\omega _{+}P}}(1-\sqrt{1-\frac{D}{P}})^{-1/2}.
\end{equation}

The ration of $E_{c.m.}$ to that for the MBO is equal to 
\begin{equation}
\mu =\left( \frac{b}{D}\right) ^{1/4}\sqrt{\frac{\omega _{+}P}{B_{1}}}(1-%
\sqrt{1-\frac{D}{P}})^{1/2}\text{.}  \label{mu}
\end{equation}

\subsection{PhO}

In a similar way, we obtain the expression (\ref{ambo}) but now with another
coefficient%
\begin{equation}
A_{pho}=\sqrt{\frac{LDB_{1}}{P}}(1-\sqrt{1-\frac{D}{P}})^{-1/2}\text{.}
\end{equation}

\section{Comparison with the Kerr metric}

In the particular case of the Kerr metric, using (\ref{k}) and concrete
values of the coefficients in (\ref{cd}), (\ref{d0}), one can obtain from
eq. (\ref{en}) - (\ref{u}) that 
\begin{equation}
\frac{E_{c.m.}}{2m}\approx \beta (2\varepsilon )^{-1/6}\text{,}  \label{be}
\end{equation}

$\beta =2^{-1/6}3^{-1/4}\sqrt{2X_{2}}$. This coincides with eq. (5.1) of 
\cite{kerr}.

Eq. (\ref{is2}) gives us%
\begin{equation}
\frac{E_{c.m.}}{m}\approx \left( \frac{2}{3}\right) ^{1/4}\sqrt{2X_{2}}%
\varepsilon ^{-1/4}\text{ }
\end{equation}%
that agrees with eq. (4.8) of \cite{kerr}. Eq. (\ref{mu}) gives $\mu
=3^{-1/4}(2-\sqrt{2)^{1/2}}$ that also agrees with the value listed in \cite%
{kerr} after eq. (5.1). One should bear in mind that the term "subcritical"
of Ref. \cite{kerr} corresponds to "usual" of our paper.

\section{Summary and conclusion}

Thus we obtained some results valid for generic dirty rotating black holes
that generalize previous observations for the Kerr metric \cite{72}, \cite%
{kerr}, \cite{gpp}. These results apply to the properties of the
near-horizon region and the BSW effect that can occur there. They include
(i) general statements about circular orbits in the near-horizon region and
the BSW effect for generic nonextremal black holes, (ii) properties of the
near-extremal black holes. In point (i), it is shown that (a) there are no
circular orbits near the horizon of a nonextremal black hole with a finite
nonzero surface gravity $\kappa $, (b) in the near-horizon region, for a
fixed small deviation of the angular momentum from the critical value, the
magnitude of the BSW effect grows from the horizon to the turning point.

If $\kappa \rightarrow 0$, so a black hole is near-extremal, circular orbits
do exist. In point (ii), for each type of a circular orbit, we showed that
their parameters depend on $\kappa $ in an universal way and investigated
two variants of the BSW effect - when one of colliding particles is moving
on such a circular orbit (O-variant) or plunges from it towards the horizon
where collides with a usual particle (H-variant). The dependence of the
collision energy $E_{c.m.}$ on $\kappa $ is found. It turned out that it has
the general character of the scaling law $E_{c.m.}\sim \kappa ^{-n}$ where $n
$ depends both on the type of the orbit and on the variant of the BSW
effect. For the ISCO, $n=\frac{1}{3}$ for the O-variant and $n=\frac{1}{2}$
for the H-variant. This difference was called "quite intriguing" in \cite%
{kerr} for the Kerr metric. Now, we see that it reveals itself for a generic
dirty rotating near-extremal black hole. For the MBO and PhO, $n=\frac{1}{2}$
in both variants. Thus the universality of black hole physics revealed
itself not only in the very existence of the BSW effect for generic dirty
black holes but also in its properties.

These results extend the potential relevance of the BSW effect for
astrophysics from the Kerr metric to dirty black hole since, say, the ISCO
is an example of how a particle's energy and angular momentum can be
fine-tuned naturally in accretion disks with electromagnetic radiation or in
inspiralling binaries \cite{kerr}.

In the present paper, our generalization of \cite{kerr} concerned the
properties of the BSW effect for particles moving in the equatorial plane.
Meanwhile, the BSW effect takes place in the Kerr background not only on the
equator but also on some finite belt around it \cite{kerr-gen}. It would be
of interest to generalize the properties of the BSW effect to nonequatorial
motion of particles near dirty rotating black holes.

\end{document}